\documentclass[aps,prd,twocolumn,superscriptaddress,nofootinbib]{revtex4-2}

\usepackage{latexsym}
\usepackage{amsmath}
\usepackage{amssymb}
\usepackage{amsfonts}
\usepackage{bm}
\RequirePackage{lineno}

\usepackage{color}
\definecolor{purple}{rgb}{0.5,0,0.5}
\definecolor{blue}{rgb}{0.0,0,0.9}
\definecolor{prdblue}{rgb}{0.133,0.118,0.498}
\usepackage[colorlinks=true, pdfstartview=FitV, linkcolor=prdblue, citecolor= prdblue, urlcolor=prdblue]{hyperref}

\usepackage{supertabular} 
\usepackage{placeins}
\usepackage{epsfig}
\usepackage{graphicx}
\usepackage{hyperref}

\hypersetup{
  breaklinks=true,
  colorlinks = true,
  linkcolor = blue,
  anchorcolor = blue,
  citecolor = blue,
  filecolor = blue,
  pagecolor = blue,
  urlcolor = blue
}
\hypersetup{
  bookmarks=true,
  bookmarksnumbered=true,
  bookmarkstype=toc,
  linktocpage=true
}

\begin{document}

\modulolinenumbers[2]
%\linenumbers

\setlength{\oddsidemargin}{-0.5cm} \addtolength{\topmargin}{15mm}

\title{\boldmath First Measurement of the $D_s^+\rightarrow K^0\mu^+\nu_{\mu}$ Decay }

\author{
  \small
M.~Ablikim$^{1}$, M.~N.~Achasov$^{4,d}$, P.~Adlarson$^{77}$, X.~C.~Ai$^{82}$, R.~Aliberti$^{36}$, A.~Amoroso$^{76A,76C}$, Q.~An$^{73,59,b}$, Y.~Bai$^{58}$, O.~Bakina$^{37}$, Y.~Ban$^{47,i}$
% ...
 \\
 \vspace{0.2cm}
 (BESIII Collaboration)\\
 \vspace{0.2cm} {\it
$^{1}$ Institute of High Energy Physics, Beijing 100049, People's Republic of China\\
% ...
$^{4}$ Budker Institute of Nuclear Physics SB RAS (BINP), Novosibirsk 630090, Russia\\
% ...
$^{36}$ Johannes Gutenberg University of Mainz, Johann-Joachim-Becher-Weg 45, D-55099 Mainz, Germany\\
$^{37}$ Joint Institute for Nuclear Research, 141980 Dubna, Moscow region, Russia\\
% ...
$^{47}$ Peking University, Beijing 100871, People's Republic of China\\
% ...
$^{58}$ Southeast University, Nanjing 211100, People's Republic of China\\
$^{59}$ State Key Laboratory of Particle Detection and Electronics, Beijing 100049, Hefei 230026, People's Republic of China\\
% ...
$^{73}$ University of Science and Technology of China, Hefei 230026, People's Republic of China\\
% ...
$^{76}$ University of Turin and INFN, (A)University of Turin, I-10125, Turin, Italy; (B)University of Eastern Piedmont, I-15121, Alessandria, Italy; (C)INFN, I-10125, Turin, Italy\\
$^{77}$ Uppsala University, Box 516, SE-75120 Uppsala, Sweden\\
% ...
$^{82}$ Zhengzhou University, Zhengzhou 450001, People's Republic of China\\    
\vspace{0.2cm}
$^{b}$ Deceased\\
% ...
$^{d}$ Also at the Novosibirsk State University, Novosibirsk, 630090, Russia\\
% ...
$^{i}$ Also at State Key Laboratory of Nuclear Physics and Technology, Peking University, Beijing 100871, People's Republic of China\\
% ...
\vspace{0.4cm}
}
}
\hspace{0.2cm}
%%%%%%%%%%%%%%%%%%%%%%%%%%%%%%%%%%%%%%%%%%%%%%%%%%%%%%%%%%%%%%%%%%%%%%%%%%%%%%%%%%%%%%%%%%
\begin{abstract}
We report the first measurement of the semileptonic decay $D^+_s \rightarrow K^0\mu^+\nu_{\mu}$, using a sample of $e^+e^-$ annihilation data corresponding to an integrated luminosity of $7.33~\mathrm{fb}^{-1}$ collected at center-of-mass energies between 4.128 to 4.226~GeV with the BESIII detector at the BEPCII collider. The branching fraction of the decay is measured to be $\mathcal{B}(D^+_s\rightarrow K^0\mu^+\nu_{\mu}) = (2.89 \pm 0.27_{\rm stat} \pm 0.12_{\rm syst})\times 10^{-3}$, where the first uncertainty is statistical and the second is systematic. Based on a simultaneous fit to the partial decay rates in $q^2$ intervals measured in $D^+_s \rightarrow K^0\mu^+\nu_{\mu}$ and $D^+_s \rightarrow K^0e^+\nu_{e}$ decays, the product value of the form factor $f^{K^0}_{+}(0)$ and the Cabibbo-Kobayashi-Maskawa matrix element $|V_{cd}|$ is measured to be $f^{K^0}_{+}(0)|V_{cd}|=0.140\pm0.008_{\rm stat}\pm0.002_{\rm syst}$. Using $|V_{cd}|=0.22486\pm0.00068$ as an input, the hadronic form factor is determined to be $f^{K^0}_{+}(0)=0.623\pm0.036_{\rm stat} \pm 0.009_{\rm syst}$ at $q^2=0$. This is the most precise determination of $f^{K^0}_{+}(0)$ in the $D^+_s \rightarrow K^0$ transition to date. The measured branching fraction and form factor presented in this work provide the most stringent test on various non-perturbative theoretical calculations. Taking $f^{K^0}_{+}(0)=0.6307\pm0.0020$ from lattice calculations as an input, we obtain $|V_{cd}|=0.220\pm0.013_{\rm stat}\pm0.003_{\rm syst}\pm0.001_{\rm LQCD}$, which is the most precise determination of $|V_{cd}|$ using the $D_s^+\rightarrow K^0\ell^+\nu_{\ell}$ decays. In addition, lepton flavor universality is tested for the first time with $D^+_s \rightarrow K^0\ell^+\nu_{\ell}$ decays in full and separate $q^2$ intervals. No obvious violation is found.
\end{abstract}

\pacs{13.30.Ce, 14.40.Lb, 14.65.Dw}% PACS, the Physics and Astronomy Classification Scheme.

\maketitle

%%%%%%%%%%%%%%%%%%%%%%%%%%%%%%%%%%%%%%%%%%%%%%%%%%%%%%%%%%%%%%%%
%%%%%     Introduction       Part                  %%%%%%%%%%%%%
%%%%%%%%%%%%%%%%%%%%%%%%%%%%%%%%%%%%%%%%%%%%%%%%%%%%%%%%%%%%%%%%
In the standard model (SM), the semileptonic (SL) decays of $D_s^+$ mesons provide valuable information on the weak and strong interactions within hadrons composed of heavy quarks~\cite{physrept494,RevModPhys67_893}. The partial decay rate is related to the product of hadronic form factor (FF) describing the strong interaction in the initial and final hadrons, and the Cabibbo-Kobayashi-Maskawa (CKM) matrix~\cite{prl10_531} element $|V_{cs(d)}|$ parametrizing the mixing between different flavors of quarks in the weak interaction. Therefore, the SL decay of $D_s^+$ mesons both allow determination of the CKM matrix element and provide rigorous tests on various nonperturbative theoretical calculations. In recent years, there has been a great deal of attention on the experimental studies of hadronic FFs in $D_s^+\rightarrow P\ell^+\nu_{\ell}$ decays~\cite{pdg24}, where $P$ and $\ell$ refer to a pseudoscalar meson and a lepton, respectively. In particular, for $D_s^+\rightarrow K^0$ transition, only $D_s^+\rightarrow K^0e^+\nu_e$ has been experimentally reported~\cite{prl122_061801,prd110_052012}.
The SL decay $D_s^+\rightarrow K^0\mu^+\nu_{\mu}$ plays an important role in understanding the $D_s^+\rightarrow K^0$ transition. However, no such measurement has been available to date. It is also found that there is roughly $2\sigma$ tensions between the $|V_{cd}|$ extracted in the $D_s^+\rightarrow K^0 e^+\nu_{e}$ decay and that in the $D\rightarrow \pi\ell\nu_{\ell}$ decays~\cite{PRD107_094516}. 
Studying the dynamics in the $D_s^+\rightarrow K^0\mu^+\nu_{\mu}$ decay will provide the necessary data to improve the precision of measuring $|V_{cd}|$ and help clarify the tension. 

Theoretical calculations on FFs in $D^+_s\rightarrow K^0$ transition are extensively carried out by a series of nonperturbative approaches~\cite{PRD62_014006,IJMPA21_6125,plb857_138975,PRD71_014020,PRD78_054002,JPG39_025005,EPJC77_587,FrontPhys14_64401,PRD98_114031,prd101_013004,prd109_026008,PRD107_094516}, which are important in understanding the quantum-chromodynamics (QCD) theory in the charm sector. However, calculations referring to the value of FF at $q^2=0$, $f^{K^0}_+(0)$, differ significantly and vary in the range from $0.57-0.82$~\cite{PRD62_014006,IJMPA21_6125,plb857_138975,PRD71_014020,PRD78_054002,JPG39_025005,EPJC77_587,FrontPhys14_64401,PRD98_114031,prd101_013004,prd109_026008,PRD107_094516}. 
In particular, a recent calculation using Lattice QCD show $f^{K^0}_+(0)=0.6307(20)$~\cite{PRD107_094516} in $D^+_s\rightarrow K^0$ transitions. 
LQCD~\cite{lattice} also predicts that FFs have minimal dependence on the spectator-quark mass, with FF values in
$D^+_s\rightarrow K^{0}$ and $D^+\rightarrow \pi^0$ transitions differing by less than 5\%.  
The verification of the LQCD FF predictions at the few percent level in SL charm decays will be helpful for further applying the
LQCD to SL $B$ decays for precise determination of CKM matrix elements $|V_{cb}|$ and $|V_{ub}|$~\cite{lattice,EPJC74_2981,prd85_114502}. 
However, the precision of available FF measurement serving for these tests is still very limited.

The SL decays of $D_s^+\rightarrow K^0\ell^+\nu_{\ell}$ also offer an excellent opportunity to test lepton flavor
universality (LFU) predicted by the SM~\cite{ARNPS,NSR}. It is suggested that the observable LFU violation effects might appear in the SL decays mediated via $c\rightarrow d\ell^+\nu_{\ell}$~\cite{CPC45_063107}. 
In recent years, a great deal of theoretical models calculated the branching fraction (BF) of $D^+_s\rightarrow K^0\mu^+\nu_{\mu}$ and the ratio $R^{\mu/e}_{K^0}=\frac{\mathcal{B}(D^+_s\rightarrow K^0\mu^+\nu_{\mu})}{\mathcal{B}(D_s^+\rightarrow K^0e^+\nu_{e})}$ between $\mu$ and $e$ modes. However, significant differences among theoretical calculations are observed with the predicted BF varying in the range $(0.20-0.39)\%$~\cite{PRD62_014006,IJMPA21_6125,PRD71_014020,PRD78_054002,JPG39_025005,EPJC77_587,FrontPhys14_64401,PRD98_114031,prd101_013004,plb857_138975}, and the $R^{\mu/e}_{K^0}$ in  $0.96-1.00$~\cite{PRD62_014006,IJMPA21_6125,PRD71_014020,PRD78_054002,JPG39_025005,EPJC77_587,FrontPhys14_64401,PRD98_114031,prd101_013004,plb857_138975}. Hence it is important to measure the BF of $D^+_s\rightarrow K^0\mu^+\nu_{\mu}$ and $R^{\mu/e}_{K^0}$ to distinguish among these theories and also test $\mu-e$ LFU.

In this Letter, we report the first study of the $D^+_s\rightarrow K^0\mu^+\nu_{\mu}$ decay, including
a measurement of the absolute BF, the determination of the FF and the CKM element $|V_{cd}|$, as well as a LFU test via $D_s^+\rightarrow K^0\ell^+\nu_{\ell}$ decays.
Throughout this paper, the charge-conjugate modes are implied unless explicitly noted.
These measurements are performed using an $e^+e^-$ annihilation data sample corresponding to an integrated luminosity of $7.33~\mathrm{fb}^{-1}$ produced at center-of-mass energies between $\sqrt{s}=4.128$ and 4.226~GeV with the BEPCII collider and collected by the BESIII detector~\cite{Ablikim:2009aa,Yu:IPAC2016-TUYA01}. 

%%%%%%%%%%%%%%%%%%%%%%%%%%%%%%%%%%%%%%%%%%%%%%%%%%%%%%%%%%%%%%%%
%%%%%     Detector and software Part               %%%%%%%%%%%%%
%%%%%%%%%%%%%%%%%%%%%%%%%%%%%%%%%%%%%%%%%%%%%%%%%%%%%%%%%%%%%%%%
Monte Carlo (MC) simulated data samples produced with a {\sc geant4}-based~\cite{geant4} software package,
which includes the geometric description of the BESIII detector and the detector response,
are used to determine detection efficiencies and to estimate backgrounds. 
The inclusive MC sample includes the production of open charm processes, the initial state radiation (ISR) 
production of vector charmonium(-like) states, and the continuum processes incorporated in {\sc kkmc}~\cite{kkmc}. 
All particle decays are modeled with {\sc evtgen}~\cite{nima462_152} using branching fractions either taken from
the Particle Data Group~\cite{pdg24}, when available, or otherwise estimated with {\sc lundcharm}~\cite{lundcharm}.
Final state radiation (FSR) from charged final state particles is incorporated using the
{\sc photos} package~\cite{plb303_163}. The generation of signal $D^+_s\rightarrow K^0\mu^+\nu_{\mu}$ incorporates knowledge of the FFs (using the $z$-series expansion model as mentioned below) obtained in this work.

\begin{figure}[tp!]
\flushleft
\begin{center}
   \includegraphics[width=\linewidth]{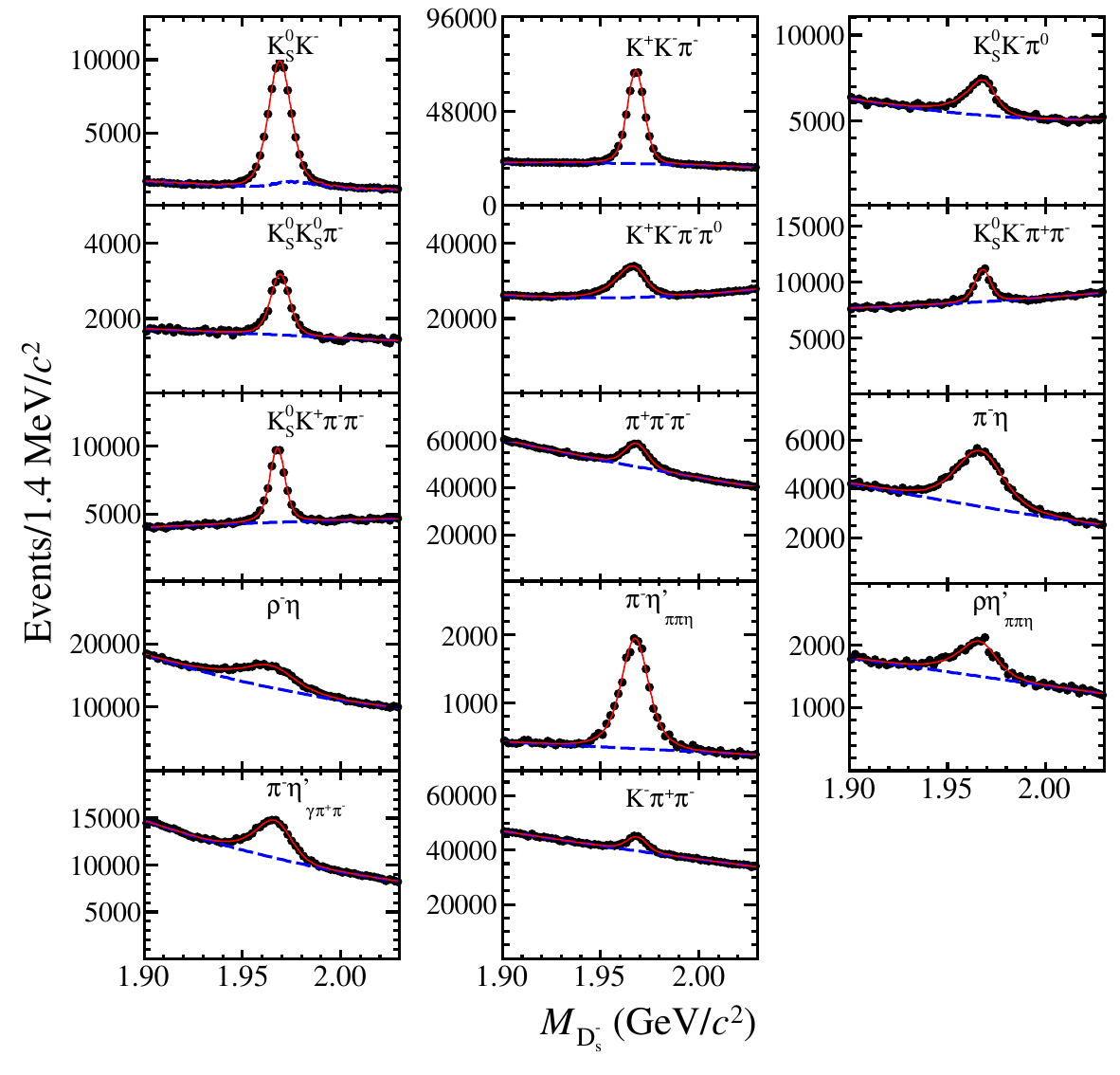}
\caption{(Color online)
Fits to $M_{D_s^-}$ distributions for the fourteen tag modes. Points with error bars are data, blue dashed curves are the fitted backgrounds and
red solid curves are the total fits.}
\label{fig:tag_mds}
\end{center}
\end{figure}

%%%%%%%%%%%%%%%%%%%%%%%%%%%%%%%%%%%%%%%%%%%%%%%%%%%%%%%%%%%%%%%%
%%%%%            Physics Analysis                  %%%%%%%%%%%%%
%%%%%%%%%%%%%%%%%%%%%%%%%%%%%%%%%%%%%%%%%%%%%%%%%%%%%%%%%%%%%%%%
The analysis presented in this analysis exploits the production of $D_s$ mesons in $e^+e^- \to D_s^{*+}D_s^- + c.c.$ and uses both ``single-tag'' (ST) and ``double-tag'' (DT) samples of $D_s$ decays~\cite{prl122_061801,prd110_052012}.
The STs are $D_s^-$ mesons reconstructed from their daughter particles in one of fourteen hadronic decays as shown in Fig.~\ref{fig:tag_mds}, while the DTs are events with a ST and a $D^+_s$ meson reconstructed as $D^+_s\rightarrow K^0\mu^+\nu_{\mu}$, with $K^0$ reconstructed via the $K^0\rightarrow K^0_S\rightarrow \pi^+\pi^-$ decay. 
The BF for the SL decay is given by~\cite{D0Kspiev,D0Kspimv}
\begin{equation}
  \mathcal{B}_{\rm SL} \,=\,
  \frac{N_{\rm DT}}{\sum_i N^{i}_{\rm ST} \,
      \left(\epsilon^i_{\rm DT}/\epsilon^i_{\rm ST}\right)} \,=\,
  \frac{N_{\rm DT}}{N_{\rm ST} \, \epsilon_{\rm SL}}, \label{eq:branch}
\end{equation}
where $N_{\rm DT}$ is the total yield of DT events, $N_{\rm ST}$ is the total ST yield, and 
$\epsilon_{\rm SL}=(\sum_i N^{i}_{\rm ST}\times\epsilon^i_{\rm DT}/\epsilon^i_{\rm ST})/\sum_i N^{i}_{\rm ST}$
is the average efficiency of reconstructing the SL decay in an ST event,
weighted by the measured yields of tag modes in the data. Here, $\epsilon^i_{\rm ST}$ and $\epsilon^i_{\rm DT}$ are the efficiencies for finding the ST and the SL decay in the 
$i$-th tag mode, respectively.

A detailed description of the selection criteria for $\pi^{\pm}$, $K^{\pm}$ and $\gamma$ candidates is given in Refs.~\cite{D0Kspiev,D0Kspimv}. 
The $\pi^0$, $\eta$, $K^0_S$, $\rho^-$, $\eta$ and $\eta^{\prime}$ mesons are reconstructed using the same technique as described in Ref.~\cite{prd110_052012}.
For all events passing the ST selection criteria, we calculate the recoil mass against the tag with the following formula
$$M_{\rm rec}=\sqrt{(\sqrt{s}-\sqrt{|\vec{p}_{D_s^-}|^2+m^2_{D_s}})^2-|\vec{p}_{D_s^-}|^2},$$
where $m_{D_s}$ and $\vec{p}_{D_s^-}$ are the
known mass~\cite{pdg24} and measured momentum of the tag $D_s^-$. In case of multiple candidates, only the candidate with $M_{\rm rec}$ closest to the known $D_s^{*+}$ mass~\cite{pdg24} is retained.
To suppress other hadronic backgrounds from non-$D_sD^*_s$ processes, requirements on the kinematic variable $M_{\rm BC} = \sqrt{(\sqrt{s}/2)^2-|\vec {p}_{D_s^-}|^2}$ are applied. The detailed requirements on $M_{\rm BC} $ are described in Ref.~\cite{prd110_052012}, and they are chosen such that the tag $D_s$ can come either directly from the production or from the $D_s^{*} \to D_s \gamma$ decay.
To determine the mode-by-mode ST yields, we perform unbinned maximum likelihood fits to the distributions
of the $D^-_s$ invariant mass $M_{D_s^-}$, as shown in Fig.~\ref{fig:tag_mds}.
Signals are modeled with the MC-derived signal shape convolved with a Gaussian function to account for
the resolution differences between data and MC, while the
combinatorial backgrounds are parameterized with second-order polynomial
functions. Due to misidentification of $\pi^-$ as $K^-$,
the backgrounds from $D^-\rightarrow K^0_S\pi^-$ form a broad peak near the $D_s^-$ known mass for $D_s^-\rightarrow K^0_SK^-$.
In the fit, the shape of this background is described using MC simulation, and its size relative to other combinatorial backgrounds is fixed.  
For each tag mode, the ST yield is obtained by integrating the signal
function over the $D^-_s$ signal region specified by $1.94<M_{D_s^-}<1.99$~GeV/$c^2$. The total ST yield summed over all fourteen ST modes is 
$N_{\rm ST}=(783.1\pm2.5)\times 10^3$, where the uncertainty is statistical only.

%%%%%%%%%%%%%%%%%%%%%%%%%%%%%
%%%%%%%%%%%%%%%%%%%%%%%%%%%%%

Candidates for the SL decay $D^+_s\rightarrow K^0\mu^+\nu_{\mu}$ are selected from the remaining tracks recoiling against the ST $D^-_s$ mesons. The $\bar{K}^0$ meson is reconstructed as a $K^0_S \to \pi^+\pi^-$ decay with the same selection criteria as on the ST side. Apart from the two charged tracks originating from $K^0_S$ decays,
it is required that there be only one additional track with charge opposite to the tagged $D_s^-$ meson. 
 For muon identification, the $dE/dx$ and TOF measurements are combined with shower properties from the electromagnetic calorimeter (EMC) to construct likelihoods for the electron, muon, and kaon hypotheses, $\mathcal{L}_e$, $\mathcal{L}_{\mu}$, and $\mathcal{L}_K$, respectively. The muon candidate must satisfy $\mathcal{L}_{\mu} > 0.001$, $\mathcal{L}_{\mu}>\mathcal{L}_e$, and $\mathcal{L}_{\mu}>\mathcal{L}_{K}$. 
Additionally, the deposited EMC energy (${\rm EMC}_{\mu}$) of the muon candidate is required to be within $0.1<{\rm EMC}_{\mu}<0.3$~GeV. The background from $D^+_s\rightarrow K^0\pi^+$ decay reconstructed as $D^+_s\rightarrow K^0\mu^+\nu_{\mu}$ is rejected by requiring the $K^0\mu^+$ invariant mass ($M_{K^0\mu^+}$) to be less than 1.70~GeV/$c^2$. Backgrounds containing additional $\pi^0$ mesons are further suppressed by requiring the maximum energy of any unused photon ($E_{\gamma \rm max}$) to be less than 0.15~GeV. 

To identify a photon produced directly from $D_s^{*\pm}$, a kinematic fit is performed by requiring
the $D^{\mp}_sD_s^{*\pm}$ pair to conserve energy and momentum in the center-of-mass frame~\cite{prl122_061801}, and the resultant $\chi^2$ is required to satisfy $\chi^2<40$.
We obtain information about the undetected neutrino with the missing-mass squared of an event, calculated by
the energies and momenta of the tag ($E_{D_s^-}$, $\vec{p}_{D_s^-}$), transition photon ($E_{\gamma}$,
$\vec{p}_{\gamma}$), and the detected SL products ($E_{\rm SL}=E_{K^{0}}+E_{\mu^+}$,
$\vec{p}_{\rm SL}=\vec{p}_{K^{0}}+\vec{p}_{\mu^+}$) by
\begin{equation}
{\rm MM}^2=(\sqrt{s}-E_{D_s^-}-E_{\gamma}-E_{\rm SL})^2-(|\vec{p}_{D_s^-}+\vec{p}_{\gamma}+\vec{p}_{\rm SL}|)^2. \nonumber
\end{equation}

\begin{figure}[tp!]
\begin{center}
   \begin{minipage}[t]{7cm}
   \includegraphics[width=\linewidth]{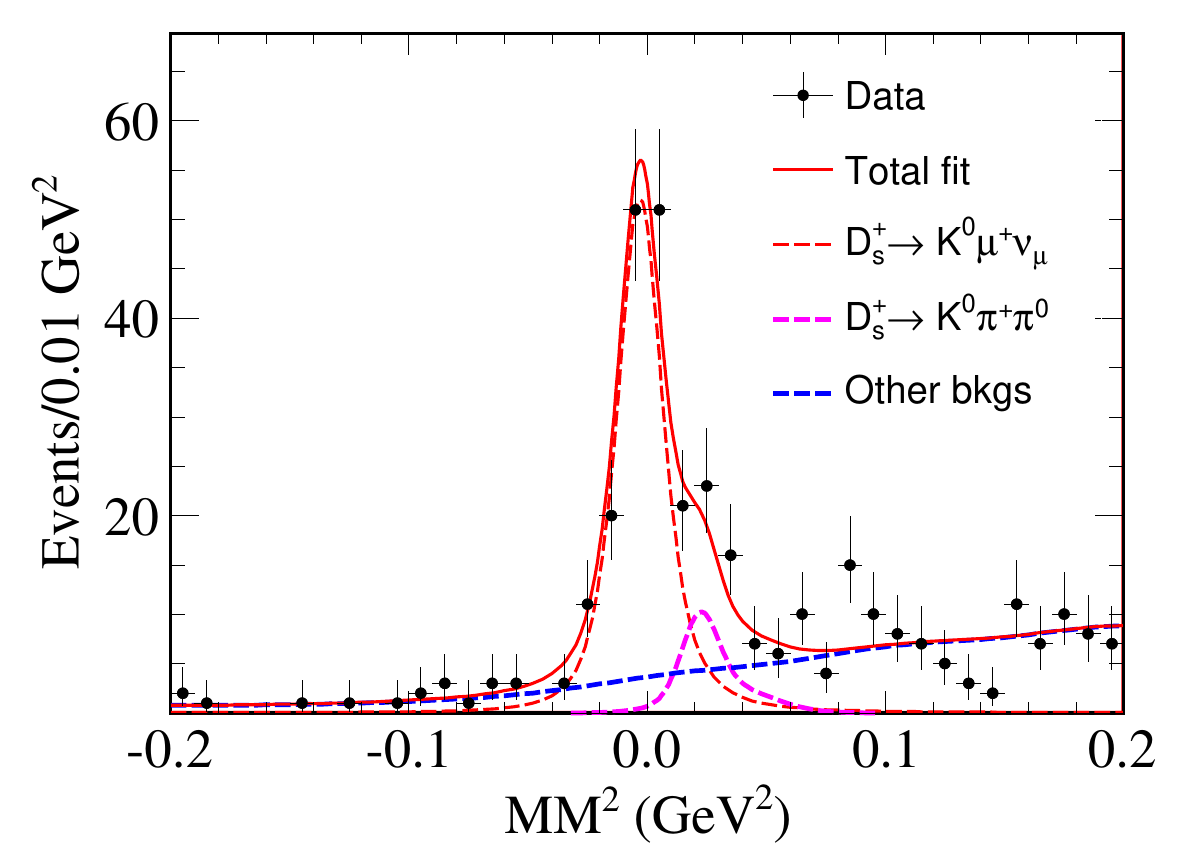}
   \end{minipage} 
   \caption{(Color online)~
   Fits to the ${\rm MM}^2$ distribution of SL candidate events.
   Dots with error bars are data, dot-dashed lines (blue) are the fitted combinatorial backgrounds
   and solid curves (red) are the total fits. The long-dashed lines (pink) show the fitted background from $D^+_s\rightarrow K^0\pi^+\pi^0$. }
\label{fig:m2miss}
\end{center}
\end{figure}

Figure~\ref{fig:m2miss} shows the ${\rm MM}^2$ distribution of the accepted candidate events for $D_s^+\rightarrow K^{0}\mu^+\nu_{\mu}$ in data.
The signal DT yield $N_{\rm DT}$ is obtained by performing an unbinned maximum likelihood fit to ${\rm MM}^2$.
In the fit, the signal is described with an MC-derived signal shape convolved with a Gaussian, 
and the combinatorial background is described by a shape obtained from the inclusive MC sample.
The residual peaking background from $D^+_s\rightarrow K^0\pi^+\pi^0$ is simulated using the MC-derived shape, with its yield fixed to 27.2 based on the MC simulation.
We obtain $N_{\rm DT}=147.1\pm13.9$ events for $D_s^+\rightarrow K^0\mu^+\nu_{\mu}$, where the uncertainties are statistical only.
No peaking backgrounds are observed in the $K^0$ mass sidebands.
The average efficiency of reconstructing the SL decay $\epsilon_{\rm SL}$ is estimated to be $(18.78\pm0.02)\%$, where the BF of $K^0\rightarrow \pi^+\pi^-$ is not included~\cite{BFK0}.  The BF of $D_s^+\rightarrow K^{0}\mu^+\nu_{\mu}$ is determined by Eq.~(\ref{eq:branch}), 
and we obtain $\mathcal B({D^+_s\rightarrow K^0 \mu^+\nu_{\mu}})=(2.89\pm0.27)\times10^{-3}$, where the uncertainty is statistical only.

%%%%%%%%%%%%%%%%%%%%%%%%%%%%%
%%%%%%%%%%%%%%%%%%%%%%%%%%%%%

With the DT technique, the BF measurements are insensitive to the systematic uncertainties of the ST selection.
The uncertainties of the $\mu^+$ tracking and PID efficiency is 1.0\% each studied with $e^+e^-\rightarrow \gamma\mu^+\mu^-$ events, while the uncertainty
due to the $K^{0}$ reconstruction is 1.5\% using the control samples $J/\psi\rightarrow K^{*\mp}K^{\pm}$ and $J/\psi\rightarrow \phi K^0K^{\pm}\pi^{\mp}$. The uncertainty associated with the ${\rm MM}^2$ fit is estimated to be 1.2\% by varying the fit ranges, 
the signal and the background shapes. The uncertainty due to the selection of the $\gamma$ is estimated to be 1.0\% based on selecting the best photon
candidate in a control sample of $e^+e^-\rightarrow D_s^{*\pm}D_s^\mp$ events with two hadronic tags, $D_s^+\rightarrow K_S^0K^+$ and $D_s^+\rightarrow K^+K^-\pi^+$.
The systematic uncertainties due to the requirements of $E_{\rm \gamma max}$ and $\chi^{2}$
are estimated using the control samples $D_s^+\to K_S^0K^+$ and $D_s^+\to K_S^0K^+\pi^0$. The differences of the acceptance efficiencies between data and MC simulation,
1.7\% and 2.0\%, are assigned as their systematic uncertainties.
The uncertainties due to the $M_{K^0\mu^+}$ requirement
is estimated to be 1.1\% by comparing the nominal BF with that measured with alternative requirement.
The uncertainty due to the MC signal modeling is estimated to be 0.9\% by varying
the input FF parameter by $\pm 1\sigma$ measured in this work.
We also consider the systematic uncertainties of $N_{\rm ST}$ (1.0\%), evaluated by using
alternative signal shapes when fitting the $M_{D_s^-}$ spectra, the quoted BF for $K_S^0\rightarrow\pi^+\pi^-$ (0.2\%), and the MC statistics (0.8\%).
The uncertainty due to different tag dependencies between data and MC simulation is estimated to be 0.2\%.
Adding these contributions in quadrature gives total systematic uncertainty of 4.2\% for
the BF measurement. Finally, we determine the BF to be $\mathcal{B}(D^+_s\rightarrow K^0 \mu^+\nu_{\mu})=(2.89\pm0.27_{\rm stat}\pm0.12_{\rm syst})\times 10^{-3}$.

%%%%%%%%%%%%%%%%%%%%%%%%%%%%%%%%%%%%%%%%%%%%%%%%%%%%%%%%%%%%%%%%
%%%%%%%%%%    formfactor       Part                %%%%%%%%%%%%%
%%%%%%%%%%%%%%%%%%%%%%%%%%%%%%%%%%%%%%%%%%%%%%%%%%%%%%%%%%%%%%%%
To improve the experimental precision of the FF parameter in the $D^+_s\rightarrow K^0$ transition, the candidates from $D_s^+\rightarrow K^0 \mu^+\nu_{\mu}$ and $D^+_s\rightarrow K^0 e^+\nu_{e}$ are analyzed simultaneously. The candidates of $D^+_s\rightarrow K^0 e^+\nu_{e}$ are selected as in Ref.~\cite{prd110_052012}. 
The partial decay rate of $D^+_s\rightarrow K^0 \ell^+\nu_{\ell}$ is measured by analyzing the $\ell^+\nu_{\ell}$ mass-squared ($q^2$) distribution. 
The $q^2$ distributions of $D^+_s\rightarrow K^0 \ell^+\nu_{\ell}$ candidates are divided into seven bins, [$m_{\ell}^2$, 0.35), [0.35, 0.70), [0.70, 1.05), [1.05, 1.40) and [1.40,~$q^2_{\rm max}$)~GeV$^2$/$c^4$, where $q^2_{\rm max}=(m_{D_s}-m_{\ell})^2$ and $m_{\ell}$ is the known mass of the lepton~\cite{pdg24}. 
The measured partial decay rate in the $i$-th $q^2$ bin, $\Delta \Gamma^i_{\rm meas}$, is determined by 
\begin{equation}
\Delta \Gamma^i_{\rm meas}=\sum\limits_{j=1}^{N_{\rm bins}}(\epsilon^{-1})_{ij}N^j_{\rm DT}/(\tau_{D_s}\times N_{\rm ST}), \nonumber
\label{eq:drate}
\end{equation}
where $\tau_{D_s}=(501.2\pm2.2)$~fs is the lifetime of $D_s^+$~\cite{pdg24} and $\epsilon_{ij}$ is an efficiency matrix to describe the reconstruction efficiency of SL decays and migration effects across $q^2$ bins~\cite{PRL124_241802,PRD101_112002}. 
The SL signal yield observed in the $j$-th $q^2$ bin, $N^j_{\rm DT}$, is obtained from a fit to the corresponding ${\rm MM}^2$ distribution. 
The measured partial decay rates $\Delta\Gamma^i_{\rm meas}$, and the differential decay rates, $\Delta\Gamma^i_{\rm meas}/\Delta q^2$, for SL decays $D_s^+\rightarrow K^0 \mu^+\nu_{\mu}$ and $D_s^+\rightarrow K^0 e^+\nu_{e}$, are summarized in Table~\ref{tab:partial}.

The expected differential decay rate in $D^+_s\rightarrow K^0 \ell^+\nu_{\ell}$  with respect to $q^2$ is expressed as~\cite{prd101_013004}
\begin{equation}\small
\frac{d\Gamma^{\rm exp}}{dq^2}=\frac{G_{F}^{2}|V_{cd}|^{2}|\vec{p}_{K^0}|^2}{96\pi^{3}m^2_{D_s}} \lambda(1-\frac{m^2_{\ell}}{q^2})^2 |f^{K^0}_+(q^2)|^2(1+\frac{m^{2}_{\ell}}{2q^{2}}),
\label{eq:dgammadq2_ksev}
\end{equation} \normalsize
where  $G_F$ is the Fermi coupling constant, $\lambda=m^4_{D_s}+m^4_{K^0}+q^4-2(m^2_{D_s}m^2_{K^0}+m^2_{K^0}q^2+m^2_{D_s}q^2)$, $|\vec{p}_{K^0}|=\lambda^{1/2}/(2m_{D_s})$ is the momentum of $K^0$ in the $D^+_s$ rest frame, $m_{K^0}$ is the known mass of $K^0$ meson~\cite{pdg24}, and the $f^{K^0}_+(q^2)$ is the hadronic FF in $D^+_s\rightarrow K^0$ transition. 

\begin{table}
\begin{center}
\caption{The measured partial decay rates for SL decays of $D_s^+\rightarrow K^0\ell^+\nu_{\ell}$ in various $q^2$ bins, where the uncertainties are statistical only~\cite{TauDs}. The systematic uncertainties of measuring 
$\Delta\Gamma^i_{K^0\mu^+\nu_{\mu}}$ ($\Delta\Gamma^i_{K^0e^+\nu_e}$) from lower to higher $q^2$ intervals are 4.2 (3.9)\%, 4.3 (4.0)\%, 4.4 (4.5)\%, 4.2 (3.8) \%, and 5.3 (6.0)\%, respectively. }
\resizebox{!}{1.7cm}{
\begin{tabular}
{l|cccc}
\hline\hline
$q^2$ interval   & $\Delta\Gamma^i_{K^0\mu^+\nu_{\mu}}$ & $\Delta\Gamma^i_{K^0e^+\nu_e}$  & $\frac{\Delta\Gamma^i_{K^0\mu^+\nu_{\mu}}}{\Delta q^2}$  & $\frac{\Delta\Gamma^i_{K^0e^+\nu_e}}{\Delta q^2}$     \\
                 &               (ns$^{-1}$)                         &                   (ns$^{-1}$)                            &                   (ns$^{-1}$GeV$^{-2}$)                       &                    (ns$^{-1}$GeV$^{-2}$) \\ 
\hline 
$[m^2_{\ell}, 0.35)$            & $1.57\pm0.31$   & $1.63\pm0.24$ & $4.62\pm0.90$ & $4.65\pm0.68$ \\
$[0.35, 0.70)$            & $1.09\pm0.24$   & $1.27\pm0.22$ & $3.12\pm0.66$ & $3.62\pm0.62$ \\
$[0.70, 1.05)$            & $0.93\pm0.23$   & $1.18\pm0.21$ & $2.65\pm0.62$ & $3.38\pm0.61$ \\
$[1.05, 1.40)$            & $1.24\pm0.25$   & $1.08\pm0.19$ & $3.55\pm0.70$ & $3.08\pm0.56$ \\
$[1.40, q^2_{\rm max})$   & $0.97\pm0.23$   & $0.84\pm0.19$ & $1.27\pm0.30$ & $1.10\pm0.24$ \\
\hline \hline
\end{tabular}
}
\label{tab:partial}
\end{center}
\end{table}
\normalsize

To extract the FF parameter, a least-$\chi^2$ fit is performed to the measured $\Delta\Gamma_{\rm meas}^i$, and the theoretically expected $\Delta\Gamma_{\rm exp}^i$, partial decay rates among $q^2$ intervals. 
The FF $f^{K^0}_+(q^2)$ in Eq.~(\ref{eq:dgammadq2_ksev}) is expressed using three theoretical parameterizations, which are the simple pole and modified pole models~\cite{plb478_417}, as well as the $z$-series expansion~\cite{SEM}. Taking into account the correlations of the measured partial decay rates among $q^2$ bins, the $\chi^2$ to be minimized is defined by
\begin{equation}
\chi^{2}=\sum_{i,j=1}^{5}(\Delta\Gamma^{i}_{\rm meas}-\Delta\Gamma^{i}_{\rm exp})C_{ij}^{-1} (\Delta\Gamma^{j}_{\rm meas}-\Delta\Gamma^{j}_{\rm exp}), 
\label{eq:chisq}
\end{equation}
where $C_{ij} = C_{ij}^{\rm stat}+C_{ij}^{\rm syst}$ is the covariance
matrix of the measured partial decay rates among $q^2$ intervals.
The statistical covariance matrix is constructed as
$C_{ij}^{\rm stat} = (\frac{1}{\tau_{D_s} \cdot N_{\rm ST}})^{2}\sum_{\alpha}\varepsilon_{i\alpha}^{-1}\varepsilon_{j\alpha}^{-1}[\sigma(N_{\rm DT}^{\alpha})]^{2}$,
where $\alpha$ labels the $q^2$ interval, and $\sigma(N_{\rm DT}^{\alpha})$ is the statistical uncertainty of DT yield observed in the $\alpha^{\rm th}$ $q^2$ interval.
The systematic covariance matrix is obtained by summing over the covariance matrix of each systematic uncertainty source, which takes the form as $C_{ij}^{\mathrm{syst}}=\delta(\Delta\Gamma^{i}_{\rm meas})\delta(\Delta\Gamma^{j}_{\rm meas})$,
where $\delta(\Delta\Gamma^{i}_{\rm meas})$ is the systematic uncertainty of the measured partial decay rate in the $i^{th}$ $q^{2}$ interval.

\begin{figure}[tp!]
\begin{center}
   \flushleft
   \includegraphics[width=\linewidth]{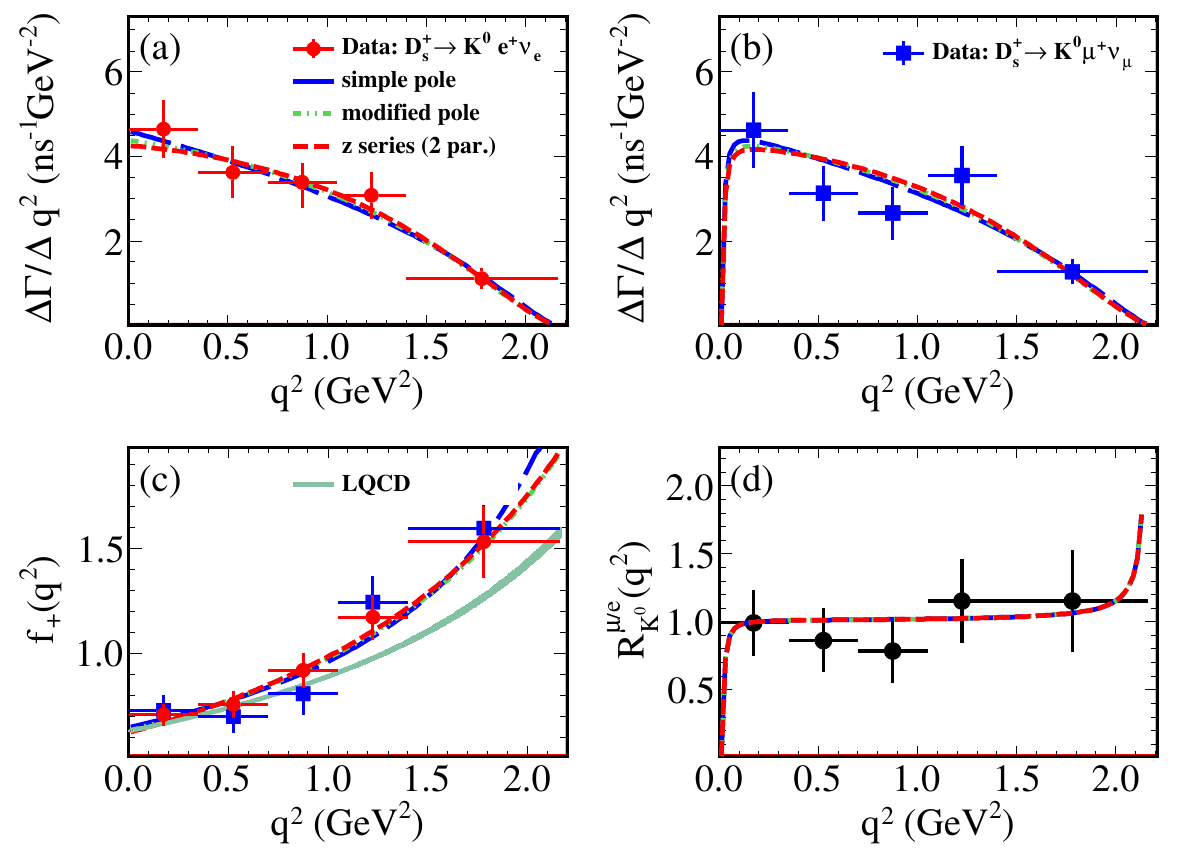}
   \caption{(Color online) Fits to the partial decay rates for (a) $D_s^+\rightarrow K^0e^+\nu_e$ and (b) $D_s^+\rightarrow K^0\mu^+\nu_{\mu}$. (c) Projection onto $f^{K^0}_+(q^2)$ for the two SL decays, while the hatched curves show the LQCD prediction in Ref.~\cite{PRD107_094516}. (d) The measured $\mathcal{R}^{\mu/e}_{K^0}(q^2)$ in various $q^2$ intervals. Dots with error bars are data, the curves show the fits with various FF parameterizations. }
\label{fig:formfactor}
\end{center}
\end{figure}

\begin{figure}[tp!]
\begin{center}
   %\flushleft
   \begin{minipage}[t]{8cm}
   \includegraphics[width=\linewidth]{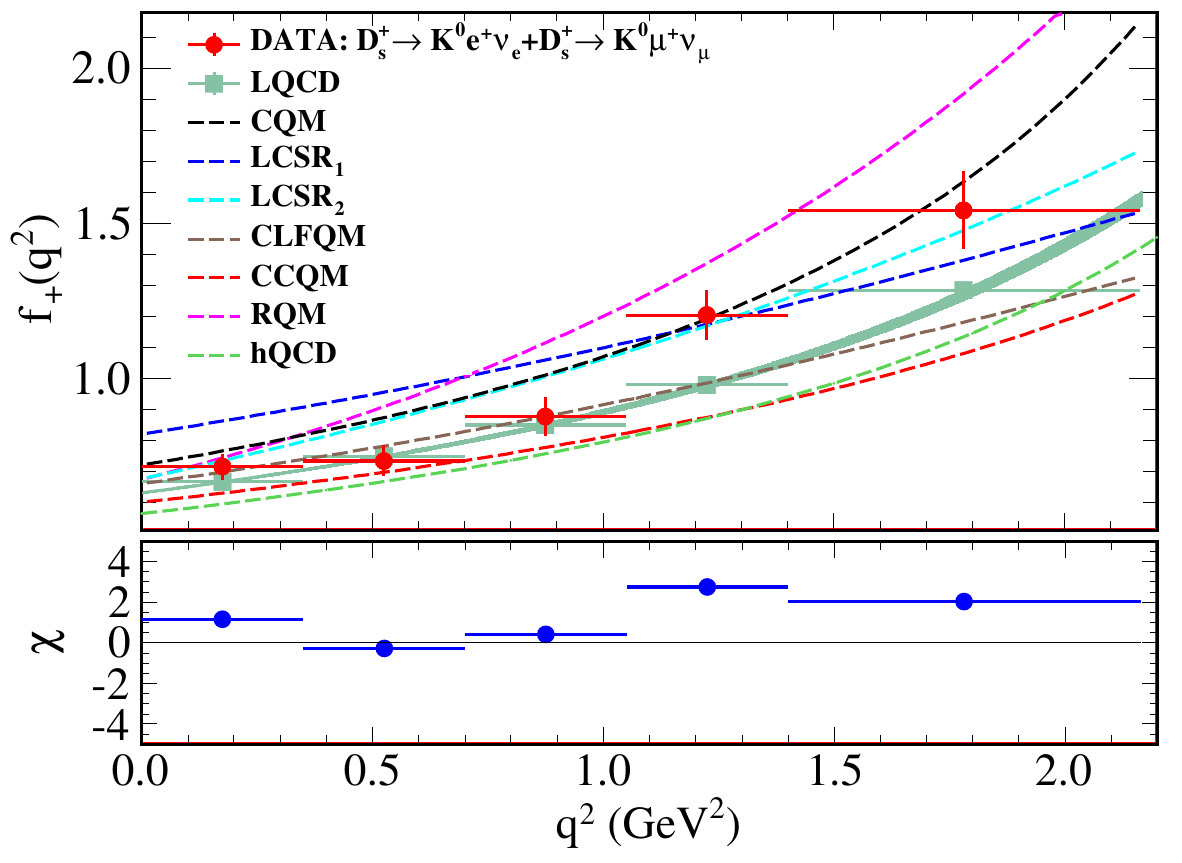}
   \end{minipage}
    \caption{(Color online) (Upper) Comparisons of $f^{K^0}_+(q^2)$ measured in this work with those calculated from the LQCD~\cite{PRD107_094516}, CQM~\cite{PRD62_014006}, LCSR$_1$~\cite{IJMPA21_6125}, LCSR$_2$~\cite{plb857_138975}, CLFQM~\cite{JPG39_025005}, CCQM~\cite{PRD98_114031}, RQM~\cite{prd101_013004}, and the hQCD~\cite{prd109_026008}. (Bottom) Distribution of $\chi=[f^{\rm meas}_+(q^2)-f^{\rm LQCD}_+(q^2)]/[\sqrt{\sigma^2_{f^{\rm meas}_+(q^2)}+\sigma^2_{f^{\rm LQCD}_+(q^2)}}]$, calculated with the measured FFs and predicted FFs of the LQCD~\cite{PRD107_094516}. }
\label{fig:cmp_formfactor}
\end{center}
\end{figure}

We perform a simultaneous fit to the partial decay rates measured in SL decays $D^+_s\rightarrow K^0 e^+\nu_e$ and $D^+_s\rightarrow K^0 \mu^+\nu_\mu$, where the two SL decays share the same parameters for the hadronic FF $f^{K^0}_+(q^2)$. Figure~\ref{fig:formfactor} show the fits to the partial decay rates and the projections of the FF $f^{K}_+(q^2)$ for $D_s^+\rightarrow K^0e^+\nu_e$ and $D_s^+\rightarrow K^0\mu^+\nu_{\mu}$ decays. The numerical results of $f^{K^0}_+(0)|V_{cd}|$ from the $\chi^2$ fit are summarized in Table~\ref{tab:sum_formfactor}. 
Taking the $|V_{cd}|=0.22487\pm0.00068$~\cite{pdg24} from the CKMfitter as an input, we also measure $f^{K^0}_+(0)$ as summarized in the last column of Table~\ref{tab:sum_formfactor}. 
The measured FF in $D_s^+\rightarrow K^0$ transition at $q^2=0$ is consistent with $0.6307(20)$ as predicted by a LQCD calculation~\cite{PRD107_094516}. 
However, the dependence of the measured FF on $q^2$ shows roughly $2\sigma$ deviations with that from the LQCD calculation in the high-$q^2$ region, as shown in Fig.~\ref{fig:cmp_formfactor}. Comparisons of the measured $f^{K^0}_+(q^2)$ with other theoretical calculations are also shown in Fig.~\ref{fig:cmp_formfactor}. 

Combining $\mathcal{B}(D^+_s\rightarrow K^0 e^+\nu_e)=(2.98\pm0.23_{\rm stat}\pm0.12_{\rm syst})\times 10^{-3}$ measured in Ref.~\cite{prd110_052012} and $\mathcal B({D^+_s\rightarrow K^{0} \mu^+\nu_{\mu}})=(2.89\pm0.27_{\rm stat}\pm0.12_{\rm syst})\times 10^{-3}$ measured in this work, we obtain the relative ratio between $\mu$ and $e$ modes to be
\begin{equation}
\mathcal{R}^{\mu/e}_{K^0}=0.97\pm0.12_{\rm stat}\pm0.04_{\rm syst}, \nonumber
\end{equation}
which is consistent with $\mathcal{R}^{\mu/e}_{K^0}=0.98099(10)^{\rm QCD}[500]^{\rm QED}$ from LQCD~\cite{PRD107_094516}, and the SM predictions in Refs.~\cite{PRD62_014006,IJMPA21_6125,PRD71_014020,PRD78_054002,JPG39_025005,EPJC77_587,FrontPhys14_64401,PRD98_114031,prd101_013004,plb857_138975}. 
In addition, we also measured the $\mathcal{R}^{\mu/e}_{K^0}(q^2)$ in various $q^2$ intervals and the result is shown in Fig.~\ref{fig:formfactor} (d), which also consistent with the SM predictions.

%%%%%%%%%%%%%%%%%%%%%%%%%%%%%
%%%%%%%%%%%%%%%%%%%%%%%%%%%%%

%%%%%%%%%%%%%%%%%%%%%%%%%%%%%
%%%%%%%%%%%%%%%%%%%%%%%%%%%%%

\begin{table*}
\begin{center}
\caption{Hadronic form factors of $D^+_s\to K^0 \ell^+\nu_{\ell}$, where the first uncertainties are statistical and the second systematic.
The correlation coefficient between the two fitted parameters is given in the fourth column.
The $\chi^2/{\rm NDOF}$ is the goodness-of-fit and NDOF is the number of degrees of freedom.}
\begin{tabular}
{lccccc}
\hline\hline
Parametrization                  & $f^{K^0}_+(0)|V_{cd}|$   &  Parameter ($M_{\rm pole}/\alpha/r_1$)  & Coefficient & $\chi^2$/NDOF & $f^{K^0}_+(0)$             \\
\hline
Simple pole~\cite{plb478_417}    & $0.144\pm0.007\pm0.001$& ~~$1.725\pm0.065\pm0.018$	&~~0.74    & $5.2/8$ & $0.640\pm0.031\pm0.005$ \\
Modified pole~\cite{plb478_417}  & $0.140\pm0.008\pm0.002$& ~~$0.636\pm0.198\pm0.056$	&$-0.83$   & $5.6/8$ & $0.623\pm0.036\pm0.009$ \\
$z$ series (two par.)~\cite{SEM} & $0.140\pm0.008\pm0.002$& $-4.000\pm1.230\pm0.346$	&~~0.87    & $5.8/8$ & $0.623\pm0.036\pm0.009$ \\
\hline \hline
\end{tabular}
\label{tab:sum_formfactor}
\end{center}
\end{table*}

%%%%%%%%%%%%%%%%%%%%%%%%%%%%%%%%%%%%%%%%%%%%%%%%%%%%%%%%%%%%%%%%
%%%%%%%%%%%%%    summary       Part                %%%%%%%%%%%%%
%%%%%%%%%%%%%%%%%%%%%%%%%%%%%%%%%%%%%%%%%%%%%%%%%%%%%%%%%%%%%%%%
%\section{Summary}
To summarize, this Letter reports the first measurement of the SL decay $D^+_s \rightarrow K^0\mu^+\nu_{\mu}$. The BF of the decay is measured to be $\mathcal{B}(D^+_s\rightarrow K^0\mu^+\nu_{\mu}) = (2.89 \pm 0.27_{\rm stat} \pm 0.12_{\rm syst})\times 10^{-3}$. 
Based on a simultaneous analysis on the decay dynamics in $D^+_s \rightarrow K^0\mu^+\nu_{\mu}$ and $D^+_s \rightarrow K^0e^+\nu_{e}$ decays, we report the most precise determination of the FF parameter in the $D_s^+\rightarrow K^0$ transition and measure $f^{K^0}_{+}(0)=0.623\pm0.036_{\rm stat} \pm 0.009_{\rm syst}$ at $q^2=0$ from the $z$-series expansion. 
In addition, taking $f^{K^0}_{+}(0)=0.6307\pm0.0020$ from the LQCD~\cite{PRD107_094516} as an input and with $f^{K^0}_+(0)|V_{cd}|$ derived from the $z$-series expansion, we also obtain $|V_{cd}|=0.220\pm0.013_{\rm stat}\pm0.003_{\rm syst}\pm0.001_{\rm LQCD}$.
The result shows a comparable precision and is consistent in $1\sigma$ with $|V_{cd}|=0.2330\pm0.0029\pm0.0133$~\cite{pdg24} from the averaged measurements via $D\rightarrow \pi\ell\nu_{\ell}$ decays.

\begin{figure}[tp!]
\begin{center}
   %\flushleft
   \begin{minipage}[t]{8cm}
   \includegraphics[width=\linewidth]{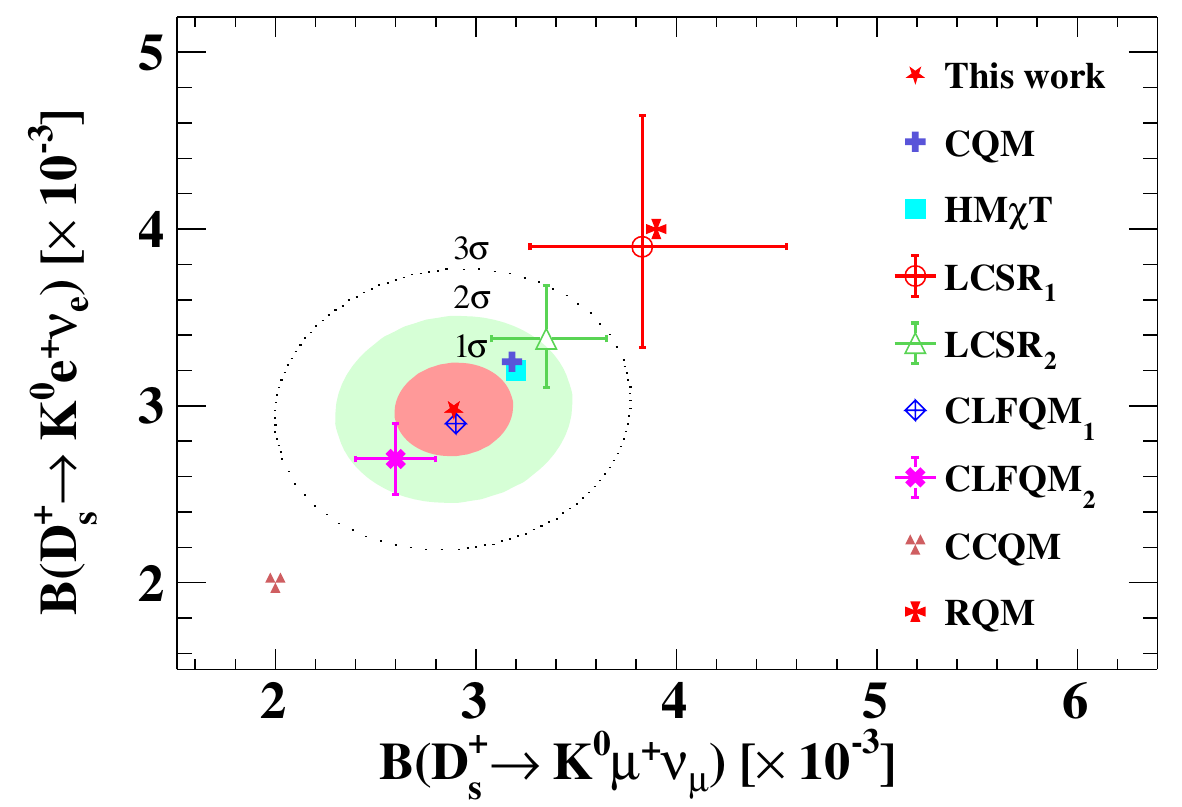}
   \end{minipage}    
   \caption{(Color online)~Comparisons of measured BFs for $D^+_s\rightarrow K^{0} e^+\nu_{e}$~\cite{prd110_052012} and $D^+_s\rightarrow K^{0}\mu^+\nu_{\mu}$ with various theoretical calculations from the CQM~\cite{PRD62_014006}, LCSR$_1$~\cite{IJMPA21_6125}, LCSR$_2$~\cite{plb857_138975}, HM$\chi$T~\cite{PRD71_014020}, CLFQM$_1$~\cite{PRD78_054002}, CLFQM$_2$~\cite{JPG39_025005,EPJC77_587}, CCQM~\cite{FrontPhys14_64401,PRD98_114031}, and RQM~\cite{prd101_013004}. The correlation coefficient between the measured BFs is 0.07. }
\label{fig:cmpbf}
\end{center}
\end{figure}

\begin{figure}[tp!]
\begin{center}
   \flushleft
   \begin{minipage}[t]{4.2cm}
   \includegraphics[width=\linewidth]{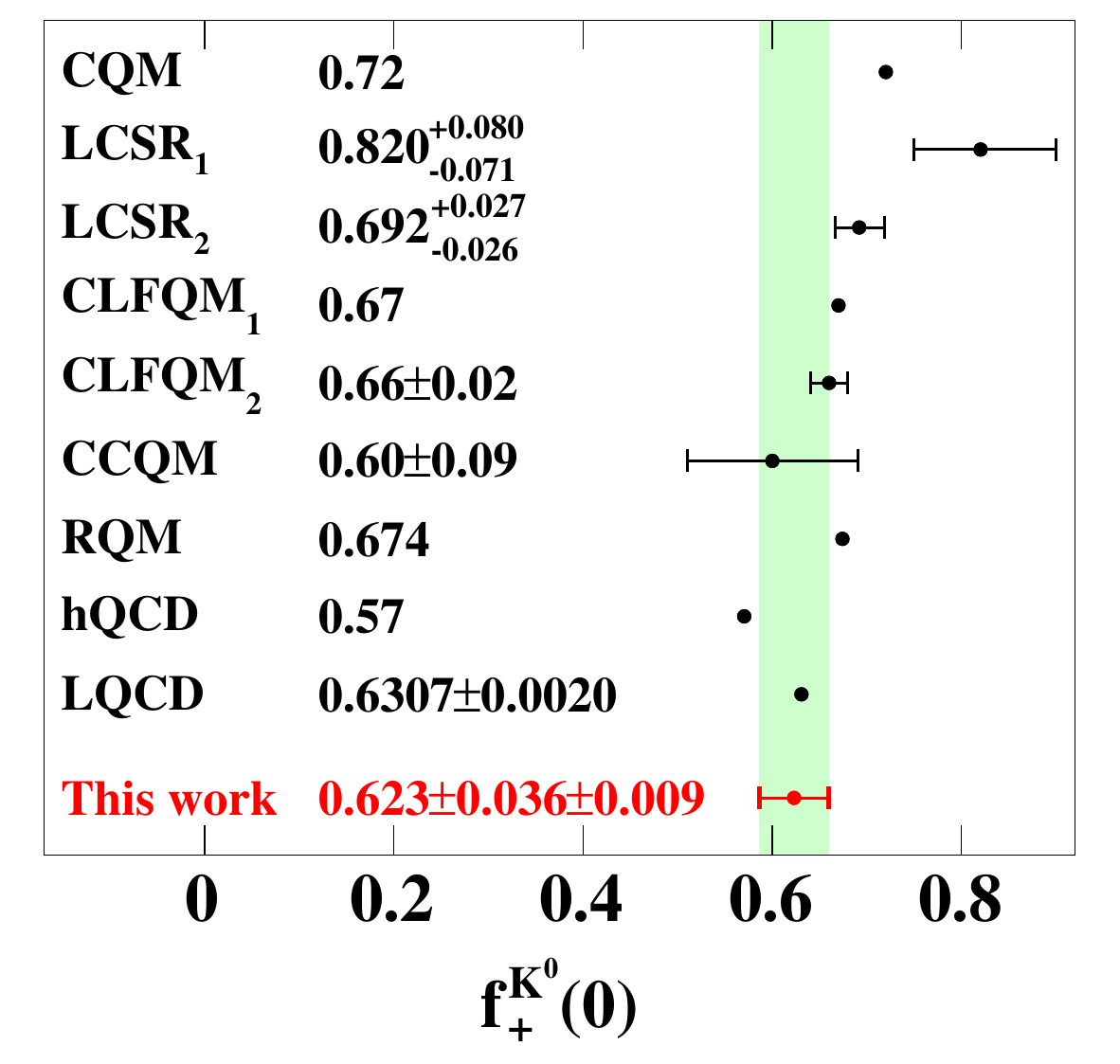}
   \end{minipage}        
   \begin{minipage}[t]{4.2cm}
   \includegraphics[width=\linewidth]{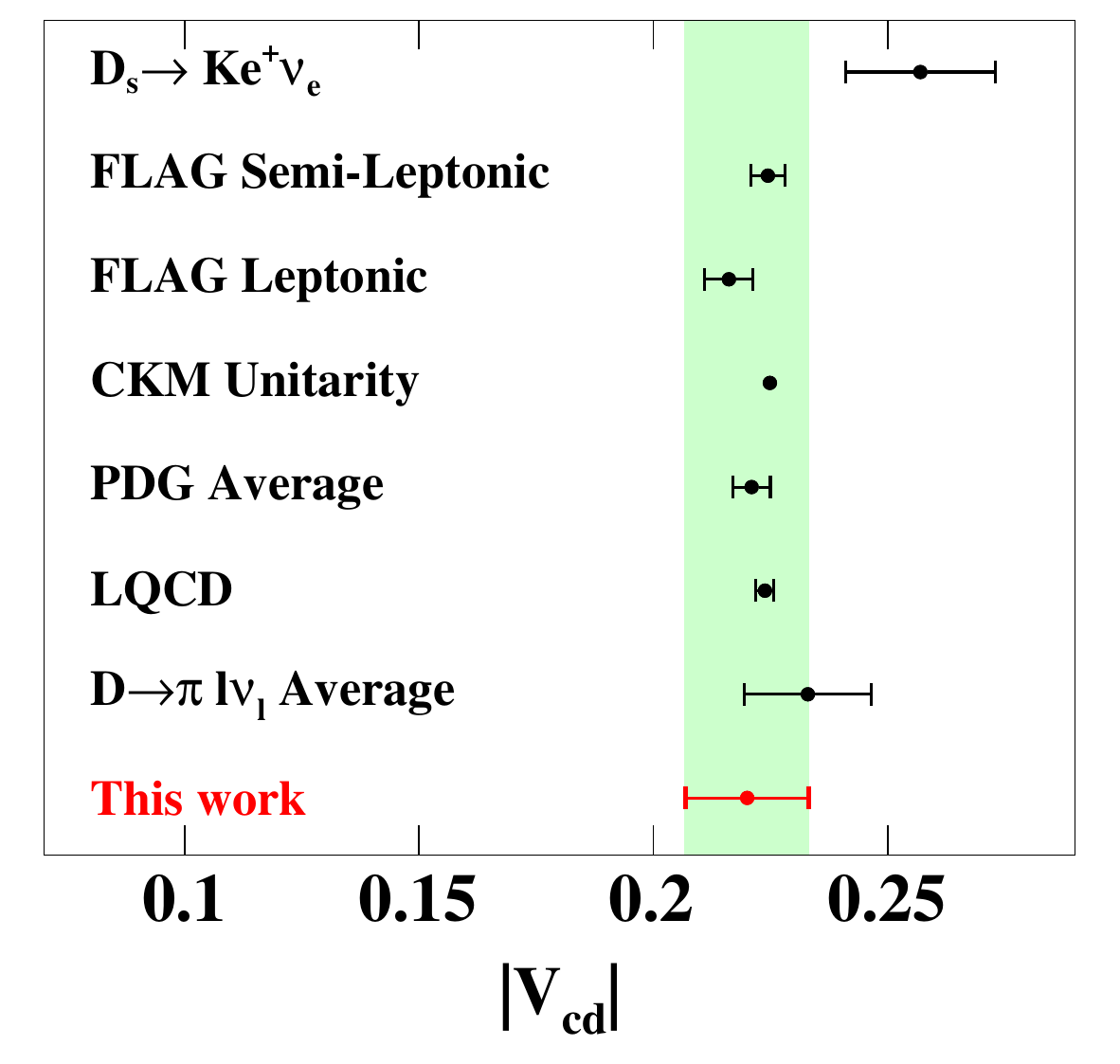}
   \end{minipage}   
   \caption{(Color online)~(Left) Comparisons of measured $f_+^{K^0}(0)$ with various theoretical calculations from the CQM~\cite{PRD62_014006}, LCSR$_1$~\cite{IJMPA21_6125}, LCSR$_2$~\cite{plb857_138975}, CLFQM$_1$~\cite{PRD78_054002}, CLFQM$_2$~\cite{JPG39_025005,EPJC77_587}, CCQM~\cite{FrontPhys14_64401,PRD98_114031}, RQM~\cite{prd101_013004}, hQCD~\cite{prd109_026008}, and the LQCD~\cite{PRD107_094516}. (Right) Comparisons of measured $|V_{cd}|$ with various results from $D_s^+\rightarrow K^0e^+\nu_e$ in Ref.~\cite{PRD107_094516}, FLAG results with $N_f=2+1+1$~\cite{FLAG}, CKM global fit~\cite{pdg24}, PDG averaged results~\cite{pdg24}, LQCD via $D\rightarrow \pi\ell\nu_{\ell}$ decays~\cite{PRD107_094516}, and averaged measurements via $D\rightarrow \pi\ell\nu_{\ell}$ decays~\cite{pdg24}.}
\label{fig:cmpff}
\end{center}
\end{figure}

The measured BF and FF presented in this work provide a stringent test on various theoretical models. Figures~\ref{fig:cmpbf} and \ref{fig:cmpff} show the comparisons of the measured BFs, FF and $|V_{cd}|$ with various results in the literature. At a confidence level of 95\%, our measured BF disfavors the central values calculated with the CCQM~\cite{FrontPhys14_64401,PRD98_114031} and RQM~\cite{prd101_013004}, and the measured FF disfavors the central value calculated with the CQM~\cite{PRD62_014006}. 
Our result for $|V_{cd}|$ rules out the $2\sigma$ tension between the values extracted via $D_s^+\rightarrow K^0 \ell^+\nu_{\ell}$ and that via $D\rightarrow \pi\ell\nu_{\ell}$ decays~\cite{PRD107_094516}. 
The LFU is also tested with $D^+_s \rightarrow K^0\ell^+\nu_{\ell}$ in full and separate $q^2$ intervals, and no LFU violation is found. 
Combining the $f^{\pi^0}_+(0)|V_{cd}|=0.1407\pm0.0025$ measured in $D^+\rightarrow \pi^0\ell\nu_{\ell}$~\cite{pdg24} and $f^{K^0}_{+}(0)|V_{cd}|=0.140\pm0.008_{\rm stat}\pm0.002_{\rm syst}$ measured in this work, we obtain $f^{K^0}_{+}(0)/f^{\pi^0}_+(0)=0.995\pm0.061$, which is consistent with LQCD prediction~\cite{lattice,EPJC74_2981,prd85_114502} and the expectation of $U$-spin ($d\leftrightarrow s$) symmetry~\cite{plb492_297}. 
These measurements provide a unique and rigorous test of the LQCD prediction that the FFs are insensitive to spectator quarks. 
However, as shown in Fig.~\ref{fig:cmp_formfactor}, our measurements for $f^{K^0}_+(q^2)$ at high-$q^2$ regions show roughly $2\sigma$ tensions with theoretical calculations from the LQCD~\cite{PRD107_094516}, CLFQM~\cite{JPG39_025005}, CCQM~\cite{PRD98_114031}, RQM~\cite{prd101_013004}, and the hQCD~\cite{prd109_026008}.
The results presented in this Letter provide stringent tests and constraints on various theoretical calculations, especially in QCD theories, and play an important role in understanding the dynamics of charm-hadron SL decays in the non-perturbative region.

%%%%%%%%%%%%%%%%%%%%%%%%%%%%%
%%%%%%%%%%%%%%%%%%%%%%%%%%%%%

%%%%%%%%%%%%%%%%%%%%%%%%%%%%%
%%%%%%%%%%%%%%%%%%%%%%%%%%%%%

%%%%%%%%%%%%%%%%%%%%%%%%%%%%%%%%%%%%%%%%%%%%%%%%%%%%%%%%%%%%%%%%
%%%%%    acknowledgments       Part                %%%%%%%%%%%%%
%%%%%%%%%%%%%%%%%%%%%%%%%%%%%%%%%%%%%%%%%%%%%%%%%%%%%%%%%%%%%%%%
\acknowledgments
The BESIII Collaboration thanks the staff of BEPCII (https://cstr.cn/31109.02.BEPC) and the IHEP computing center for their strong support. This work is supported in part by National Key R\&D Program of China under Contracts Nos. 2023YFA1606000, 2023YFA1606704; National Natural Science Foundation of China (NSFC) under Contracts Nos. 11635010, 11935015, 11935016, 11935018, 12022510, 12025502, 12035009, 12035013, 12061131003, 12192260, 12192261, 12192262, 12192263, 12192264, 12192265, 12221005, 12225509, 12235017, 12361141819, 12375090; the Chinese Academy of Sciences (CAS) Large-Scale Scientific Facility Program; CAS under Contract No. YSBR-101; 100 Talents Program of CAS; The Institute of Nuclear and Particle Physics (INPAC) and Shanghai Key Laboratory for Particle Physics and Cosmology; 
ERC under Contract No. 758462; German Research Foundation DFG under Contract No. FOR5327; Istituto Nazionale di Fisica Nucleare, Italy; Knut and Alice Wallenberg Foundation under Contracts Nos. 2021.0174, 2021.0299; Ministry of Development of Turkey under Contract No. DPT2006K-120470; National Research Foundation of Korea under Contract No. NRF-2022R1A2C1092335; National Science and Technology fund of Mongolia; Polish National Science Centre under Contract No. 2024/53/B/ST2/00975; STFC (United Kingdom); Swedish Research Council under Contract No. 2019.04595; U. S. Department of Energy under Contract No. DE-FG02-05ER41374. This paper is also supported by the Fundamental Research Funds for the Central Universities, and the Research Funds of Renmin University of China under Contract No. 24XNKJ05.

%%%%%%%%%%%%%%%%%%%%%%%%%%%%%%%%%%%%%%%%%%%%%%%%%%%%%%%%%%%%%%%%
%%%%%    bibliographies       Part                %%%%%%%%%%%%%
%%%%%%%%%%%%%%%%%%%%%%%%%%%%%%%%%%%%%%%%%%%%%%%%%%%%%%%%%%%%%%%%


\begin{thebibliography}{99}
\bibitem{BFK0} The BFs of $K^0\rightarrow K^0_S$ and $K^0_S\rightarrow \pi^+\pi^-$ are not included in the detection efficiency $\epsilon_{\rm DT}$. 
The $K^0_S$ fraction of a $K^0$ is taken as 50\% and the BF of $K^0_S\rightarrow \pi^+\pi^-$ is taken from PDG~\cite{pdg24}.
\end{thebibliography}
\end{document}